# The Challenge of Cell Segmentation in Spatially Resolved Transcriptomics

Naveed Ishaque[†,1], Peter Kharchenko[†,2], Daria Lazic[†,3], Jieran Sun[†,4,5], Jean Yee Hwa Yang[†,6], Martin Emons[7,8], Florian Heyl[9,10,11], Wolfgang Huber[3], Daniel Jones[12], Louis B. Kuemmerle[13], Alex R. Lederer[4], Malte D. Luecken[13,14,15], Vinicius Maracaja-Coutinho[16,17,18], Matthias Meyer-Bender[3], Andrew Moorman[19], Evan W. Newell[12,20], Quan Nguyen[21,22], Shyam Prabhakar[23], John Randell[24], Daria Romanovskaia[13,14,15], Oliver Stegle[9,3], Gary D. Bader[‡,25], Raphael Gottardo[‡,4,5]

† These authors contributed equally

‡ Co-corresponding authors

Affiliations

[1] Berlin Institute of Health at Charité – Universitätsmedizin Berlin, Center of Digital Health, Berlin, Germany

[2] Harvard Medical School, Boston, MA, USA

[3] Genome Biology Unit, European Molecular Biology Laboratory (EMBL), Heidelberg, Germany

[4] Biomedical Data Science Center, Lausanne University Hospital (CHUV), Lausanne, Switzerland

[5] University of Lausanne, Lausanne, Switzerland

[6] The University of Sydney, Sydney, NSW, Australia

[7] Department of Molecular Life Sciences, University of Zürich, Winterthurerstrasse 190, 8057 Zürich, Switzerland

[8] Swiss Institute of Bioinformatics, University of Zurich, Winterthurerstrasse 190, 8057 Zürich, Switzerland

[9] Division of Computational Genomics and Systems Genetics, German Cancer Research Center (DKFZ), Im Neuenheimer Feld 280, 69120 Heidelberg, Germany

[10] The German Human Genome-Phenome Archive (GHGA), Im Neuenheimer Feld 280, 69120 Heidelberg, Germany

[11] Centro Nacional de Análisis Genómico (CNAG), Barcelona, Spain

[12] Fred Hutchinson Cancer Center, Seattle, WA, USA

[13] Institute of Computational Biology, Helmholtz Munich, Neuherberg, Germany

[14] Institute of Lung Health and Immunity (LHI), Helmholtz Munich, Comprehensive Pneumology Center (CPC-M), Neuherberg, Germany

[15] German Center for Lung Research (DZL), Germany

[16] Unidad de Genómica Avanzada (UGA), Advanced Center for Chronic Diseases (ACCDiS), Facultad de Ciencias Químicas y Farmacéuticas, Universidad de Chile, Santiago, Chile

[17] Laboratório de Medicina e Saúde Pública de Precisão, Instituto Gonçalo Moniz, Fundação Oswaldo Cruz, Fiocruz, Salvador, Brazil

[18] Instituto Nacional de Ciência e Tecnologia em Saúde Digital (INCT-DigiSaúde), Salvador, Brazil

[19] Single-Cell Analytics Innovation Lab, Memorial Sloan Kettering Cancer Center, New York, NY, USA

[20] National University of Singapore, Singapore

[21] QIMR Berghofer Medical Research Institute, Brisbane, Queensland, Australia

[22] Institute for Molecular Biosciences, The University of Queensland, St. Lucia, Queensland, Australia

[23] Genome Institute of Singapore, Singapore

[24] Human Cell Atlas, Cambridge, MA, USA

[25] The Donnelly Centre, University of Toronto, Toronto, ON, Canada

## Summary

Spatially resolved transcriptomics (SRT) is transforming how we study tissues by measuring gene expression in cells in their spatial context. However, the field lacks robust methodological guidance on one of its most fundamental analytical steps: how to accurately segment cells and assign spatially localized transcripts to them. Major technical challenges include sparse molecular signals, transcript displacement, complex cellular morphologies, and the projection of three-dimensional tissue architecture onto two-dimensional imaging planes. These challenges make segmentation a major source of uncertainty, with errors that can propagate through downstream analyses and ultimately lead to misleading biological interpretations. Here, we argue that segmentation should be treated as a central unresolved problem in spatial omics rather than a routine preprocessing step. We review current approaches, highlight key methodological limitations, including the lack of appropriate metrics and gold-standard benchmarks, and propose a community-driven path forward. Establishing shared evaluation frameworks, scalable benchmark datasets, and transparent reporting standards will be essential for transforming SRT into a robust and reproducible foundation for biological discovery and clinical translation.

## Introduction

Single-cell and single-nuclei RNA sequencing (sc/snRNA-seq) have transformed biology by enabling comprehensive cell states characterization through gene expression profiling at the level of individual cells (Siletti *et al.*, 2023). Continued technological advances and decreasing costs, together with global initiatives such as the Human Cell Atlas (Regev *et al.*, 2017), have democratized these approaches and resulted in the generation and integration of extensive organ-scale datasets, resulting in so-called cell atlases. These resources can be readily queried by the scientific community to investigate targeted biological hypotheses. However, scRNA-seq (which we use from here on to also encompass snRNA-seq) requires tissue dissociation, which results in a loss of spatial information. Since gene expression in cells of multicellular organisms unfolds in the context of the spatial architecture of tissues, where neighboring cells can interact, organize into niches, and collectively regulate biological function, scRNA-seq is limited in the study of how cells influence each other *in situ*.

Recent technological advances for spatially resolved transcriptomics (SRT) greatly aid in quantifying gene expression within intact tissues. Imaging-based platforms such as Xenium (Janesick *et al.*, 2023), MERSCOPE (Chen *et al.*, 2015), and CosMx (He *et al.*, 2022), and modern sequencing-based platforms such as Stereo-seq (Chen *et al.*, 2022), Visium HD (Oliveira *et al.*, 2025), and Open-ST (Schott *et al.*, 2024) can provide cellular and even subcellular resolution (**Figure 1A**). These technologies have already uncovered major biological insights and hold promise for advancing both basic research and clinical applications (Gong *et al.*, 2024; X. Wang *et al.*, 2024; Seydel, 2025; Liu, Dai and Wang, 2026).

Despite these advances, a central methodological challenge remains: accurate cell segmentation, defined as the precise delineation of individual cellular entities - often inferred from morphological boundaries - and the correct assignment of transcripts to their cells of origin (**Box 1**). While traditionally treated as sequential steps, recent approaches address cell delineation and transcript assignment as a single, joint optimization problem, which we refer to as transcript-informed segmentation. For simplicity, we will use the term segmentation throughout and specify transcript-informed segmentation when clarification is needed. Robust and accurate segmentation is essential for quantifying gene expression variability within cells across tissues, integration of data across samples, and alignment of spatial measurements with single-cell reference atlases (Yu *et al.*, 2025). Segmentation errors can produce mixed or spurious cellular expression profiles, directly biasing cell-type annotations and distorting inferred cell-cell interactions (Bilous *et al.*, 2026; Mitchel *et al.*, 2026). Although cell segmentation has long been recognized in the broader field of image analysis as a critical step for extracting biological insight from microscopy data (Ma *et al.*, 2024), the unique technical constraints of SRT – including the lack of universal membrane markers, the sparsity of transcript detections, physical and optical displacement effects on transcripts, and the projection of complex 3D tissue architecture into 2D measurements (**Figure 1B**, **Box 2**) – make segmentation particularly challenging. Despite substantial progress over the past two years, driven by improved tissue staining strategies and advances in artificial intelligence (AI)-enabled image-based cell segmentation, the core limitations of the problem remain largely unsolved.

There are more than a dozen segmentation-related methods for SRT, each using different sources of information and adopting its own modelling approaches (Carpenter *et al.*, 2006; Pachitariu, Rariden and Stringer, 2025); (Sawada and Honda, 2009; Schmidt *et al.*, 2018; Littman *et al.*, 2020; Petukhov *et al.*, 2022; Fu *et al.*, 2024; Heidari *et al.*, 2025); (Qian *et al.*, 2020; Prabhakaran, 2022; Si *et al.*, 2024; Müller-Bötticher *et al.*, 2025). It remains unclear whether any particular class of methods consistently outperforms others across diverse technologies and tissue types. This uncertainty is driven in part by the absence of gold-standard references and segmentation-specific evaluation metrics. Most likely, while several algorithms perform well in specific settings, none are yet broadly adequate, and because SRT data is expensive to collect and growing rapidly, there is an urgent need for systematic and coordinated methodological development and evaluation. Our goal is to clarify and refine what cell segmentation should entail in the context of SRT, discuss limitations of existing approaches, and issue an open call to the community for shared metrics that are suitable for evaluating SRT cell segmentation, coordinated benchmarking efforts, and new datasets to enable rigorous evaluation and advance the field.

## The Cell Segmentation Challenge and Why it's important

### A shift in the segmentation paradigm

In traditional tissue-imaging modalities such as histological and fluorescence microscopy, cell segmentation has been a relatively well-defined task. Historically, it focused on delineating cell boundaries, including membranes and nuclei, to match morphological contours, a paradigm originating from early digital image analysis and classical computer-vision methods (Meijering, 2012; Y. Wang *et al.*, 2024). Within this morphology-driven framework, segmentation quality is typically assessed by how accurately predicted masks align with visible cellular structures.

Segmentation methods developed within this paradigm are often applied to SRT; however, the underlying signals differ fundamentally and the analytical objectives of these methods are more limited. In SRT, the central object is no longer simply the cell boundary, but the reconstruction of an accurate cell-resolved molecular profile from spatially-distributed transcripts. Membrane and cell interior stains (e.g. CD45, E-cadherin, VIM, 18s rRNA) generate spatially continuous intensity fields that approximate cellular morphology. In contrast, SRT provides additional informative modalities, most notably discrete transcript molecule measurements. These transcript-derived signals are count-based, yielding sparse point patterns in imaging-based platforms and sparse, lattice-based count fields in sequencing-based assays. Additional complexity arises from technical and biological factors (**Box 2**). Primary mechanisms such as transcript diffusion and optical localization errors can physically or computationally shift inferred transcript positions (Ståhl *et al.*, 2016; Linares *et al.*, 2023; You *et al.*, 2024). Non-cellular RNA, arising from sources such as ambient RNA and severe cell truncation due to tissue sectioning, may further contribute to spatially displaced signals (Marco Salas *et al.*, 2025). These processes manifest as data-level phenomena such as spillover or spatial bleeding and ultimately result in contamination or admixture (*i.e.,* mixed expression profiles that cannot be attributed to a single cell)(Mitchel, Gao, *et al.*, 2025; Bilous *et al.*, 2026). Together, they blur the distinction between technical artifacts and biologically meaningful variation, underscoring that accurate transcript-to-cell assignment is a central challenge in SRT (**Figure 1B**).

Nonetheless, morphology remains essential. Many SRT platforms rely heavily on immunofluorescence(IF)-based membrane or nuclear staining, H&E staining, or a combination of both, with manual visual inspection remaining crucial for interpreting tissue architecture and morphometric properties (Ozirmak Lermi *et al.*, 2025). However, even a perfectly delineated cell boundary does not guarantee correct transcript assignment, as transcripts within that boundary may originate from neighboring cells, background, or from cells located in adjacent z-planes (**Box 1**) that project into the same focal plane (Tiesmeyer *et al.*, 2026). Although imaging-based SRT technologies can detect overlapping cells when sufficient images along the z-axis are acquired, sequencing-based methods without z-localization cannot readily correct projection-induced ambiguities. Consequently, transcript-agnostic segmentation approaches developed for morphology-based imaging and 2D boundary delineation cannot be directly applied without modification, and accurate inference of cell boundaries from transcript-resolved data remains inherently challenging.

**Segmentation errors can severely distort downstream inference**

In low-plex assays, which typically measure tens of markers or fewer, segmentation primarily supports relatively straightforward downstream analyses, such as per-cell quantification of mRNA or protein abundance, morphological characterization (e.g., shape, texture, polarity), and limited marker-based phenotyping. For these assays, although segmentation remains challenging, the downstream analytical tasks are comparatively simpler and primarily involve cell counting or the assignment of single-marker intensities to cells, resulting mainly in localized quantitative biases (de Souza, Zhao and Bodenmiller, 2024). In contrast, in high-plex spatial omics, profiling hundreds to thousands of transcripts, each segmented cell represents a high-dimensional molecular profile that underpins clustering, spatial modeling, niche inference, and cell-cell interaction analyses. As a result, segmentation inaccuracies producing contamination or admixture can give rise to systematic biases, yielding artificial cell states, spurious cell-cell signaling inferences, and distorted spatial organization (**Figure 1C**, **Table 1**). Thus, segmentation errors do not merely introduce random noise; they can create reproducible but misleading molecular patterns that are easily mistaken for biological signals. These effects are particularly severe when neighboring cell types differ markedly in total RNA content, allowing highly transcriptionally active cells to dominate local transcript signals. For example, malignant tumor cells may contain an order of magnitude more mRNA than adjacent immune cells, such that even minor segmentation errors can systematically bias downstream biological interpretation (Kwok *et al.*, 2025; Bilous *et al.*, 2026).

# Current Solutions and Approaches

### Transcript-agnostic and multichannel segmentation

Early pipelines for image-based cell segmentation employed classical computer-vision techniques such as intensity thresholding, watershed transforms, active contours, and seeded expansion, and these are still in use. For instance, the nucleus-centered expansion approach used by numerous vendors can be viewed as region-based segmentation, where nuclear boundaries act as seeds based on which cell contours are approximated. Some manufacturers, including 10x Genomics (Xenium In Situ Multimodal Cell Segmentation), Resolve BioSciences (Molecular Cartography (MC2, 2025)), Bruker (CosMx SMI Human Universal Cell Segmentation Kit), and Vizgen (MERSCOPE Cell Boundary Kit), offer commercial multichannel segmentation support incorporating DAPI, membrane, or cell interior stains to mitigate the limitations of nuclear-based expansion; however, nuclear-centric segmentation remains the default when membrane markers are absent. Other vendors describe how users can perform their own immunofluorescence experiments to support transcript-agnostic segmentation (Pachitariu and Stringer, 2022; Pachitariu, Rariden and Stringer, 2025)

### AI for transcript-agnostic segmentation

Recent progress in AI has substantially improved the accuracy, generalizability, and robustness of cell segmentation. Convolutional neural network architectures such as U-Net (Ronneberger, Fischer and Brox, 2015), Mask R-CNN (He *et al.*, 2017), Cellpose (Stringer *et al.*, 2021), and StarDist (Schmidt *et al.*, 2018), and more recently, transformer-based foundation models including SAM (Segment Anything Model) (Kirillov *et al.*, 2023) and variants fine-tuned for microscopy (Archit *et al.*, 2025), now enable versatile, data-driven segmentation that better captures heterogeneous cell morphology and tissue context (**Figure 1D).** These advances are used in commercial SRT platforms, such as Bruker's CosMx and Vizgen's MERSCOPE, which both employ Cellpose-based segmentation on multichannel immunofluorescence images (Ozirmak Lermi *et al.*, 2025). These models are highly effective for H&E-based nuclei segmentation, which is often used to generate cell masks in sequencing-derived SRT, such as Visium HD (Oliveira *et al.*, 2025), Stereo-Seq (Chen *et al.*, 2022), high-definition spatial transcriptomics (HDST) (Vickovic *et al.*, 2019), and Seq-Scope (Anacleto *et al.*, 2026). Beyond strict single-cell analyses, such masks can also be used to improve the effective resolution of spot-based data using methods like Spotiphy (J. Yang *et al.*, 2025), HEDeST (Gortana *et al.*, 2026), and PanoSpace (He *et al.*, 2026).

### Transcript-informed segmentation

Several algorithms tailored to imaging-based SRT have emerged that directly use transcript-level information to improve segmentation (**Figure 1D)**, including Baysor (Petukhov *et al.*, 2022), BIDcell (Fu *et al.*, 2024), ComSeg (Defard *et al.*, 2024), Segger (Heidari *et al.*, 2025), and ProSeg (Jones *et al.*, 2025). These methods combine morphological cues with SRT distributions, and in some cases extend to full three-dimensional coordinates to refine transcript assignment at single-cell resolution. For example, ProSeg, Baysor, and Segger explicitly model the 3D spatial organization of individual transcripts, enabling more accurate delineation of cell boundaries and signal separation than is achievable with purely 2D approaches. However, while transcript-based segmentation offers advantages over morphology-driven approaches using imaging data, optical artifacts (**Box 2**) and transcript sensitivity can impair transcript detection accuracy, making careful panel design essential for reliable cell boundary inference.

### Transcript-informed segmentation correction

Complementing these segmentation frameworks are transcript correction methods that operate conditional on a given cell mask to reduce background signal, mitigate transcript spillover, and improve the accuracy of cell-level expression estimates (**Figure 1D)**. These approaches employ heuristic and probabilistic frameworks to separate cell-intrinsic expression from ambient contamination, leveraging ideally high-purity cell-type expression profiles obtained from scRNA-seq as reference signatures. Probabilistic cell typing by in situ sequencing (pciSeq) (Qian *et al.*, 2020) exemplifies this approach, using scRNA-seq-derived cell type gene expression profile references to infer both cell boundaries and type identities directly from in situ transcript detection. resolVI (Ergen and Yosef, 2025) and MisTIC (Y. Yang *et al.*, 2025) use variational autoencoders to estimate and subtract spillover from neighboring cells. SPLIT (Bilous *et al.*, 2026) uses reference single-cell RNAseq data to perform cell-type deconvolution to separate primary from secondary expression signatures, removing the latter to produce cleaner cellular profiles. FastReSeg (Wu, Beechem and Danaher, 2024) introduces an approach that scores each transcript for its goodness-of-fit within its assigned cell using log-likelihood ratios, flags likely segmentation errors, and selectively reassigns mislocalized transcripts. CellAdmix (Mitchel *et al.*, 2026) uses factorization of molecular neighborhood graphs to detect and remove from the data likely admixture molecules. Ovrlpy (Tiesmeyer *et al.*, 2026) partitions the transcript distribution into upper and lower z-plane subsets and identifies regions where gene expression differs between the two, suggestive of vertically overlapping cells.

## Gaps with current approaches?

### Limits of transcript-agnostic segmentation

Despite the paradigm shift in segmentation driven by SRT technologies, segmentation quality is still often evaluated primarily through visual inspection of cell morphological boundaries. While morphology-based assessments remain valuable for interpretability and can help guide transcript-to-cell assignment, segmentation in SRT should not be viewed as a purely morphological task, as one of the primary objectives of spatial transcriptomics is the accurate reconstruction of cell-level gene expression profiles. In this context, cell morphology-driven approaches can miss cells with strong transcript signal, a limitation that is particularly pronounced for H&E-based segmentation (H&E staining provides indirect and often ambiguous cellular boundaries) and for cell types whose molecular identity is poorly reflected by standard morphological markers (e.g., erythrocytes, which lack a nucleus and are therefore poorly captured by DAPI-based segmentation). Increasing evidence (Jones *et al.*, 2025) suggests that the most accurate strategies integrate structural cues with transcript-level information, moving toward transcript-informed segmentation frameworks that jointly model cellular morphology and molecular signal.

### Impacts of segmentation errors on cellular and spatial inference

Even minor segmentation errors can lead to significant downstream consequences (Mitchel, Gao, *et al.*, 2025; Bilous *et al.*, 2026) (**Table 1, Figure 1C**). For example, spatial contamination, i.e. the misassignment of transcripts across cell boundaries, can substantially alter inferred cell-cell signaling inference and lead to divergent biological conclusions, underscoring the need for caution in interpretation (Mitchel, Gao, *et al.*, 2025; Bilous *et al.*, 2026). Although some segmentation and correction tools partially mitigate these errors, they do not fully resolve the issue of spatial contamination. Disentangling biological variation from technical artifacts in cell-cell RNA spatial patterns remains an open problem (Salas *et al.*, 2025; Bilous *et al.*, 2026)

### Limitations of single-cell RNA-seq as a reference

Whether used for cell gene expression profile purification from mixed signals or segmentation quality assessment, approaches that use scRNA-seq data as a reference often implicitly assume it provides a ground-truth representation of cell-intrinsic expression. However, such reference datasets are frequently derived from different samples and may not faithfully represent the corresponding spatial data, due to biological (e.g. different sets of cells sampled) and technical differences (e.g. gene coverage, panel design, and sensitivity). Moreover, scRNA-seq data are subject to their own technical and biological limitations, including missing or overrepresented cell populations, dissociation effects, the presence of doublets or empty droplets, and platform-specific biases (Heumos *et al.*, 2023). Finally, while scRNA-seq methods assess an average profile per cell or nucleus, high-resolution SRT can capture more intricate patterns within a cell that are not captured by typical scRNA-seq methods, such as nuclear or cytoplasm compartmentalization, or transcript polarization (Mitchel *et al.*, 2026).

**Spatial data needs methods that scale**

SRT technologies enable profiling of cells across large tissue sections and routinely generate datasets that are an order of magnitude larger than dissociation-based single-cell experiments, and this is before considering data at the transcript-level. SRT data generate dense gene expression matrices due to higher molecular detection sensitivity, restricting the use of efficient data structures, such as sparse matrix representations, and related algorithms. As throughput continues to scale, the community must prepare for datasets containing billions of spatially resolved cells, each with tens of thousands of individual transcript points. Meeting this demand will require advances in data structures, compression, distributed computing, and scalable analytical frameworks (Atta and Fan, 2021; Fang *et al.*, 2023). Emerging community standards such as OME-Zarr (Moore *et al.*, 2023) for cloud-native, chunked storage of large multidimensional bioimaging data, together with integrative spatial data infrastructures such as the SpatialData (Marconato *et al.*, 2025) framework in Python and MoleculeExperiment (Peters Couto *et al.*, 2023) or SpatialFeatureExperiment (Moses *et al.*, 2023) in R, will be essential for managing (multimodal) spatial omics data at scale. Their continued development, interoperability, and long-term maintenance will be critical to ensure reproducible and scalable analysis as dataset sizes grow.

**Reproducibility and robustness remain poorly characterized**

Overall, there is a lack of systematic assessment of reproducibility, sensitivity, and robustness for current segmentation methods in SRT. In many cases, it remains unclear how sensitive algorithmic outputs are to changes in input data, model initialization, parameter choices, or how segmentation performance is impacted by other preprocessing steps (e.g., normalization, cell type calling). Similarly, reproducibility across biologically comparable samples, such as adjacent tissue sections or technical replicates, has rarely been formally assessed. As a result, users are often left without clear guidance on how to evaluate segmentation quality in the absence of appropriate ground truth (including both data and metrics), limiting confidence in downstream biological conclusions.

# Metrics, Benchmarks, and Evaluation

**Segmentation-specific metrics are needed**

Metrics are critical for assessing and benchmarking computational methods, including cell segmentation. Although some groups have proposed metrics for evaluating spatial transcriptomics platforms and analytical pipelines (Plummer *et al.*, 2025; Totty, Hicks and Guo, 2025), these efforts generally place little emphasis on the downstream impact of segmentation performance. This reflects a deeper challenge: segmentation errors arise from multiple sources (e.g. irregular cell morphology, transcript diffusion, optical artifacts, and projecting 3D to 2D), and no single quantitative readout can comprehensively summarize these diverse failure modes.

In the absence of a ground truth, evaluations of current segmentation methods have generally relied either on consistency of alternative morphological evidence or transcriptional state. As an example of the first category, in evaluating Baysor (Petukhov *et al.*, 2022), the authors compared predicted segmentations to independent segmentation masks derived from poly(A) signal (or an outer membrane immunostaining signal when it was not used for the initial segmentation). While this allowed authors to compare the performance of different segmentation methods, the resulting precision and recall estimates are inherently relative as no individual modality is able to detect all the cell boundaries accurately, and all are limited by the accuracy and availability of the orthogonal staining data (Petukhov *et al.*, 2022). The second category uses transcriptional plausibility metrics. For example, in evaluating Proseg (Jones *et al.*, 2025), the authors derived “spurious” gene pairs from changes in coexpression under nuclear boundary expansion and scored methods by relative spurious coexpression. In evaluating BIDCell (Fu *et al.*, 2024), the authors quantified agreement with scRNA-seq references and marker structure using Pearson correlations of average expression profiles and marker-based purity from curated markers to derive precision-recall summaries. However, the metrics rely on scRNA-seq data and suffer from the limitations discussed above.

Other characteristics, such as the number of transcripts per cell, cell size distributions, or the total cell counts that could be used as metrics, are difficult to interpret. For example, oversegmentation may artificially inflate cell counts, while undersegmentation may inflate transcript count per cell. Such failure modes are often indistinguishable under commonly used summary statistics. As a result, reliable interpretation typically requires external reference points, such as using the same characteristics from other imaging modalities or scRNA-seq platforms (e.g. total number of transcripts per cell from matched samples) to assess segmentation quality.

These limitations highlight the need for interpretable metrics that assess (i) cell boundary delineation accuracy, and (ii) transcript-to-cell assignment accuracy (**Box 1**). Additionally, evaluation frameworks should include both reference-based metrics (e.g., based on scRNA-seq or spatial proteomics) and reference-free metrics (*e.g.* batch integration or cluster separation), and should ideally be able to disentangle true biological signal (e.g., cell-cell signalling) from technical contamination arising from misalignment or transcript spillover (**Box 2**).

**The field needs better gold standards**

While expert knowledge and extensive training data support morphology-based assessment, benchmarking transcript-level precision is far more difficult. This difficulty arises in part from the lack of definitive ground truth describing how gene expression varies across spatial contexts, distinct microenvironments, and even within individual cellular compartments. Attempts to generate crowdsourced, human-labeled ground-truth data have been proposed (Luecken *et al.*, 2025). However, constructing manual ground truth at the level of individual cells and transcripts is both practically infeasible and fundamentally ill-defined, as such ground truth does not exist (we don't know the true expression level of genes in individual cells). As a result, several alternative approaches have been suggested, such as the generation of experimentally controlled data using tools that transfer dissociated cells onto spatial slides to image thousands of genes in intact cells (e.g., STAMP (Pitino *et al.*, 2025)) or sequence on the slides (e.g., Stereo-Cell (Liao *et al.*, 2025)). Because these approaches operate on intact cells arranged in a monolayer rather than cross-sectioned tissue, they can provide ground-truth segments under simplified conditions. However, they currently don't capture the more complex features of real SRT data, such as cells that overlap in 3D (Sui *et al.*, 2025).

Simulation-based frameworks offer another promising direction, enabling systematic control over transcript distributions, cell overlap, and noise characteristics (Qian *et al.*, 2024; Yang *et al.*, 2026). While such simulations can be valuable for method development and methodological validation, they rely on modeling assumptions and are therefore not expected to fully capture the complexity of real tissues and tend to favor methods with similar assumptions.

Still, external data can provide valuable complementary metrics for evaluation. To date, most approaches have relied on scRNA-seq reference datasets, now widely available through initiatives such as the Human Cell Atlas (Regev *et al.*, 2017), to assess concordance between scRNA-seq profiles and spatially reconstructed gene expression. These references have been used, for example, to transfer cell-type annotations in order to evaluate cell-type separation, assess gene sensitivity, or identify biologically implausible co-expression patterns. For the latter, metrics such as the Mutually Exclusive Co-expression Rate (MECR; (Plummer *et al.*, 2025)) and Negative Marker Purity (Fu *et al.*, 2024; Marco Salas *et al.*, 2025) provide sensitive, biologically grounded readouts of contamination in segmentation outputs, enabling robust comparisons across methods. Notably, these metrics operate at the level most affected by segmentation errors - cell-type identity and gene expression fidelity. However, even the highest-quality single-cell atlases are derived from dissociated tissues, which disrupt native spatial organization and may introduce transcriptional perturbations during dissociation that could cause them to differ compared to SRT measurements of the same sample.

The closest approximation to a single, highly sensitive performance metric is achieved when ground-truth information is available within the same dataset, most notably through multiplexed protein staining. Membrane protein stains can assist in delineating cell boundaries, particularly in tissues with elongated or irregular cell morphologies, while surface markers support cell phenotype identification and enable direct assessment of concordance between protein and RNA expression. In such settings, comparisons between predicted and annotated boundaries or phenotypes can reveal performance differences that transcript-based or scRNA-seq-derived metrics may fail to capture. However, spatial proteomics is itself subject to technical noise, and no single marker panel can comprehensively label all cell types. Even high-quality membrane stains often fail to reliably resolve cell boundaries across diverse tissues (Ozirmak Lermi *et al.*, 2025), leading to inconsistent or ambiguous cell labels and contours. Moreover, transcript misassignment due to diffusion and optical artifacts (Ståhl *et al.*, 2016; Linares *et al.*, 2023; You *et al.*, 2024; Marco Salas *et al.*, 2025) or 3D into 2D projection (Tiesmeyer *et al.*, 2026) weakens the correspondence between RNA and protein data. Collectively, these biological and technical constraints make exhaustive, annotation-based benchmarking inherently incomplete.

**Current benchmarks are limited**

The field currently lacks standardized, unbiased benchmarks for segmentation in spatial omics. Most comparisons are confined to individual method papers and rely on bespoke datasets, limited evaluation metrics, and selectively chosen baselines. Only a small number of independent evaluations exist, and these typically span restricted technology types and datasets (Marco Salas *et al.*, 2025; Plummer *et al.*, 2025; Yu *et al.*, 2025; Bilous *et al.*, 2026) and limited tissue diversity (Luecken *et al.*, 2025), thereby constraining generalizability. Moreover, the time required to develop, release, and publish benchmarks often causes them to become outdated by the time they are available. Living benchmarking frameworks, such as Open Problems in Single-Cell Analysis (Luecken *et al.*, 2025), BenchHub (Liang *et al.*, 2025), OpenEBench (Capella-Gutierrez *et al.*, 2017), and Omnibenchmark (Mallona *et al.*, 2024), offer a promising approach to mitigate this issue by enabling continuous updates from the scientific community as new methods and datasets emerge.

A further obstacle is the challenge of hyperparameter selection. Many segmentation algorithms require iterative tuning that interacts with tissue type, staining quality, imaging depth, and transcript sparsity. This complicates fair comparisons as performance differences may reflect tuning efforts rather than intrinsic algorithmic capability, as has been observed in other fields (Musgrave, Belongie and Lim, 2020). A shared generalized segmentation framework with systematic, automated, or at least user-guided hyperparameter optimization would enable more equitable benchmarking across methods and technologies.

## Emerging Directions

### Towards multimodal segmentation

Although advances in experimental technologies, such as improved chemistries and reduced transcript displacement, may ease some aspects of cell segmentation, the shift toward richer multimodal readouts and full 3D profiling is likely to increase analytical complexity and make cell segmentation more challenging overall (Sui *et al.*, 2025). At the same time, this growing complexity brings new opportunities, for example, with additional modalities, such as high-plex protein staining, expanding the information available for validation or moving towards transcript-informed segmentation. Many platforms now support multimodal panels that combine high-plex RNA and protein measurements (He *et al.*, 2021), allowing protein signals to inform transcript-to-cell assignment and to help enforce concordance across molecular layers.

### Cell-segmentation-free approaches

While cell segmentation will likely remain essential for cell-level inference and atlas-scale efforts, cell-segmentation-free approaches, such as Spage2Vec (Partel and Wählby, 2021), Baysor-NCV (Petukhov *et al.*, 2022), SSAM (Park *et al.*, 2021), FICTURE (Si *et al.*, 2024), Points2Regions (Andersson *et al.*, 2024), TopAct (Benjamin *et al.*, 2024), STHD (Sun and Zhang, 2025) or Sainsc (Müller-Bötticher *et al.*, 2025), that operate directly on individual transcript coordinates offer an intriguing alternative (**Figure 1D)**. These methods model transcript distributions without predefined cell boundaries and can be applied in either supervised or unsupervised settings to group transcripts into putative spatial domains. In this framework, spatial transcriptional coherence rather than cell morphological continuity serves as the primary objective. A major advantage of these approaches is that they do not require predefined cell boundaries and generate interpretable spatial maps of molecular identity. These spatial maps can then be classified to produce cell-type maps, analogous to semantic rather than instance segmentation. As such, they allow rapid assessment of data quality prior to cell segmentation, and provide a reference framework to identify biologically meaningful cell type signals that segmentation-based approaches may fail to capture. However, a major tradeoff with cell-segmentation-free approaches is that since they do not delineate individual cells, they confine the scope of the downstream analysis to questions that do not explicitly involve cell-level measurements, thus rendering cell-segmentation-free approaches complementary to segmentation-based approaches.

**AI-driven, segmentation-free image analysis**

A conceptually related class of approaches includes AI models that operate directly on images, such as vision transformers and their recent multimodal extensions (Liu *et al.*, 2023; Ding *et al.*, 2025; Kuehl *et al.*, 2025; Shaban *et al.*, 2025), which extract biological information directly from image tiles without explicit cell segmentation. These models provide another form of segmentation-free analysis, positioned between transcript-level modeling and cell-aggregate representations, and have demonstrated impressive performance for H&E data in particular. However, they also face major limitations in biological interpretability, as the features driving model predictions are often difficult to map back to specific cellular or molecular mechanisms. While strict interpretability may not be essential for diagnostic applications, it is likely to play a much more critical role in target discovery and drug development.

**Modeling subcellular structure**

Subcellular transcript localisation to organelles (e.g., nucleus, cytoplasm) represents an underexplored dimension of SRT analysis. APEX-seq (Fazal *et al.*, 2019) maps compartment-specific transcriptomes by proximity labelling within defined organelles, while Bento (Mah *et al.*, 2024) leverages single-molecule coordinates to annotate transcript localization domains (e.g., nuclear edge, cell periphery) and quantify gene colocalization, together offering existing frameworks for a bottom-up, organelle-scaffolded approach to cell modelling in SRT.

**Advancing biological and clinical insight**

Improved segmentation will accelerate biological discovery in both fundamental research and clinical applications. This impact will be particularly pronounced for large-scale SRT efforts, which aim to characterize cellular organization, behavior, and their links to pathogenesis within native tissue contexts. At the same time, emerging SRT datasets, combined with new modeling approaches and complementary single-cell–level measurements from external modalities (e.g., fluorescence microscopy), will enable a more refined understanding of intracellular and extracellular transcriptional processes. These insights will improve the accurate modeling of the expected abundance and spatial distribution of transcripts within and around cells, which in turn will further enhance segmentation accuracy.

# Outlook and Call to Action

### Opportunities and risks of large-scale data generation

As large-scale spatial datasets such as MOSAIC (MOSAIC Consortium and Hoffmann, 2025), MANIFEST (Lim *et al.*, 2025), HCA (Regev *et al.*, 2017), HuBMAP (Turner *et al.*, 2025), and PHAROSAI (*PharosAI*, no date) continue to emerge, they offer unprecedented opportunities to train more accurate segmentation frameworks; however, if model output segmentation quality is poor, the same data volumes can amplify errors and propagate misleading biological signals at scale. Many of these frameworks aim to train foundation models (Theodoris *et al.*, 2023; Cui *et al.*, 2024; Ding *et al.*, 2026) to learn generalizable representations that capture cell identity and organization within tissue context. However, model generalizability will depend not only on scale but also on the quality, diversity, and representativeness of the underlying training data. Recent discussions on AI-powered precision medicine have emphasized that biased or underrepresented single-cell datasets can embed inequities into computational models, particularly when models are applied to populations or disease contexts absent from training data (Maracaja-Coutinho and Nakaya, 2026). Without rigorous quality control and robust transcript-informed segmentation, these models risk internalizing segmentation artefacts and spatial biases rather than true biological structure (**Figure 2**). This risk is compounded when datasets are not sufficiently representative across tissues, technologies, ancestries, environments, and disease contexts, reinforcing the need for inclusive, well-annotated, AI-ready spatial omics resources. It is therefore imperative for the field to develop and access improved segmentation methods, keep pace with state-of-the-art advances, and promote open sharing of both raw and processed data as well as ground truth.

### Community-driven and sustainable benchmarking frameworks

Taken together, the limitations described throughout this paper underscore the need for open, community-driven benchmarking efforts that explicitly address both the absence of a true gold standard and the fragmented landscape of evaluation metrics. An ideal benchmark would integrate curated datasets spanning diverse tissues, technologies, and staining conditions, and incorporate orthogonal and ideally three-dimensional, morphological or multimodal references that serve as proxies for ground truth. Achieving such breadth and diversity in benchmarking data requires coordinated community efforts and shared resources, such as Open Problems (Luecken *et al.*, 2025), BEACON (Ren *et al.*, 2024), Omnibenchmark (Mallona *et al.*, 2024), BenchHub (Liang *et al.*, 2025), and GESTALT (Plummer, Vlachos and Martelotto, 2025). These datasets could be generated by individual laboratories, through large-scale data generation initiatives, or by technology developers providing robust baseline resources for community-wide evaluation. To maximize their utility, all data should be standardized wherever possible and shared alongside the necessary raw and associated inputs (e.g., individual z-stacks, H&E images, protein stains), as well as comprehensive metadata (e.g., versioning, resolution, acquisition parameters). In addition, derived outputs should be distributed in widely adopted, interoperable formats (e.g., HDF5, Zarr, AnnData, SummarizedExperiment) to facilitate benchmarking, drawing on established FAIR practices and lessons learned from initiatives such as the Human Cell Atlas.

Accessible and reproducible educational infrastructures will also be essential to ensure broad community participation in benchmarking efforts and method adoption, particularly as spatial omics analysis becomes increasingly computationally complex. Open notebook-based ecosystems and community-driven training frameworks can help democratize access to advanced analytical workflows, facilitate transparent method comparison, and promote reproducible computational practices across diverse research environments (Possik *et al.*, 2025; Rojas-Hidalgo *et al.*, 2026).

Equally important, the value of such shared benchmark datasets depends on how they are evaluated. Benchmarking should therefore rely on clearly defined, biologically grounded complementary metrics that capture distinct and orthogonal aspects of segmentation performance, rather than a single aggregate score. These metrics should be combined using principled, transparent multi-metric scoring strategies that make trade-offs explicit and interpretable. To remain relevant as methods and technologies evolve (**Figure 2**), both the datasets and evaluation procedures must be embedded within a living, updatable framework. Such a framework would also enable consistent and transparent reporting of methodological choices, including tuning strategies and hyperparameter optimization, ensuring fair, reproducible, and meaningful comparisons across segmentation approaches.

Only through coordinated community action (**Figure 2**) - spanning data generation, metric definition, evaluation practices, method development, and the long-term stewardship of benchmarking resources - can the field advance toward SRT segmentation standards that match the scale and ambition of the technology. In the absence of such coordination, suboptimal segmentation and substantial variation in segmentation approaches will persist. Because segmentation is an upstream step for virtually all downstream analyses, this will propagate errors throughout analytical pipelines, leading to increased variability, poor comparability across studies, and ultimately a reproducibility crisis. Conversely, a robust community-driven benchmarking ecosystem could not only accelerate progress in segmentation itself but also establish shared standards, interoperable tools, and collaborative practices that strengthen the broader spatial biology ecosystem. Such infrastructure could further serve as a foundation for addressing other major computational challenges in the field, including complex problems such as cell–cell communication inference and three-dimensional tissue reconstruction.

## Acknowledgements

This work was supported by the Chan Zuckerberg Initiative via grant CZIF2022-007488 - HCA Data Ecosystem (to MDL). JS and RG were supported by the Swiss National Science Foundation(SNSF) grant 320030 215550. JYHY is supported by the Australian National Health and Medical Research Council (NHMRC) Investigator Grant APP2017023. SP is supported by the Agency for Science Technology and Research, Singapore under the Industry Alignment Fund - Pre-Positioning (IAF-PP) grant H24J4a0001 and the A*STAR-AMED Joint Grant R23I2IR072. QHN is supported by the Australian Research Council (ARC DECRA Grant DE190100116), the National Health and Medical Research Council (NHMRC Project Grant 2001514), and the NHMRC Investigator Grant (GNT2008928). FH was funded by the German Federal Ministry of Research, Technology and Space (BMFTR; 01KD2206A,B,L,M,/SATURN3). ME was funded by a travel supplement to Swiss National Science Foundation (SNSF) grant 204869. WH and MMB were supported through the European Union's Horizon 2020 Research and Innovation Programme under grant agreement No. 964264 (BIALYMP) in subprojects 01KT2311A and 01KT2311B. VMC was supported by the Agencia Nacional de Investigación y Desarrollo (ANID, Chile) through the FONDAP (15130011 and 1523A0008) and Anillo (ATE220016) programs. In addition, VMC is a recipient of a Research Productivity Fellowship from the Conselho Nacional de Desenvolvimento Científico e Tecnológico (CNPq, Brazil; grant 314606/2026-2) and received funding through the DigiSaúde-INCT initiative (grant 408775/2024-6). DR and MDL are supported by funding from the European Union's Horizon 2023 HORIZON MISS CANCER-01-01 programme under Grant Agreement No. 101136552 (SPACETIME).

### Declaration of generative AI and AI-assisted technologies in the manuscript preparation process

During the preparation of this work the author(s) used ChatGPT to assist with proofreading the manuscript. After using this tool/service, the author(s) reviewed and edited the content as needed and take(s) full responsibility for the content of the published article.

### Declaration of Interests

N.I.'s laboratory has received research grants (funding to the institution) from Owkin.

M.D.L. consults for CatalYm GmbH.

R.G. has received consulting income (payments made to Lausanne University Hospital) from Takeda, Arcellx, Sanofi, and Owkin, and declares ownership in Ozette Technologies. R.G. has received research funding from Owkin and 10x Genomics.

All other authors declare no competing interests.

## Figure 1. Schematic overview of the current spatially resolved transcriptomics (SRT) segmentation landscape.

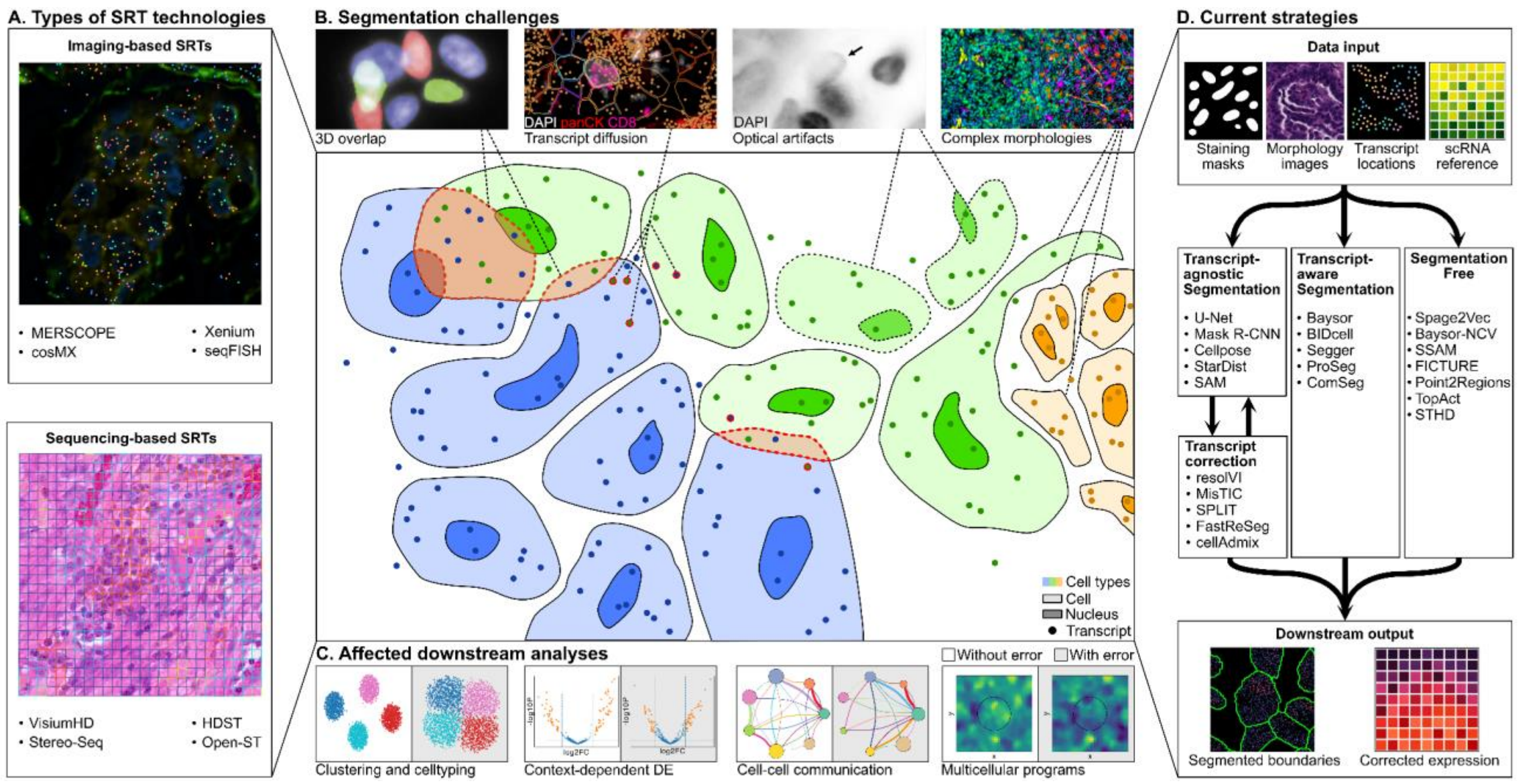

**Figure 1 (A)** Examples of different SRT technologies used as input for segmentation methods. Example images generated from publicly available spatial transcriptomics datasets from 10x Genomics, including the Xenium FFPE human breast biomarkers dataset (*Xenium v1 Human Breast FFPE with Biomarkers & Housekeeping Genes Custom Panel*, no date); top) and the Visium HD FFPE human colorectal cancer dataset (*Visium HD Spatial Gene Expression Library, Human Colorectal Cancer (FFPE)*, no date); bottom). **(B)** Top: Selected common challenges encountered when segmenting SRT datasets (**Box 2**); Bottom: illustrative drawing for different types of segmentation challenges, with cell types indicated by different colors. Image credits: "3D overlap" adapted from (Molnar *et al.*, 2016); "Transcription diffusion" and "Optical artifacts" adapted from (Bilous *et al.*, 2026); "Complex morphologies" adapted from Watson and colleagues (Watson *et al.*, 2024). **(C)** Downstream analyses potentially affected by segmentation quality with exemplary visualizations, including, from left to right, UMAPs for four cell types, volcano plots of differential expression between two regions, circle plots for cell-type communication, and spatial plots of niche enrichment scores (**Table 1**). **(D)** Current strategies for segmenting SRT datasets, with representative exemplary methods discussed in the main text (**Box 1**). Image credits: **"**Morphology images" adapted from (Bilous *et al.*, 2026); "Segmentation boundaries" adapted from (Wu, Beechem and Danaher, 2025).

**Acronyms:** SRT, spatially resolved transcriptomics; DE, differential expression; scRNA, single-cell RNA sequencing; H&E, hematoxylin and eosin; UMAP, Uniform Manifold Approximation and Projection.

**Figure 2. Current gaps and call to action for sustained progress in spatial transcriptomics–driven cell segmentation.**

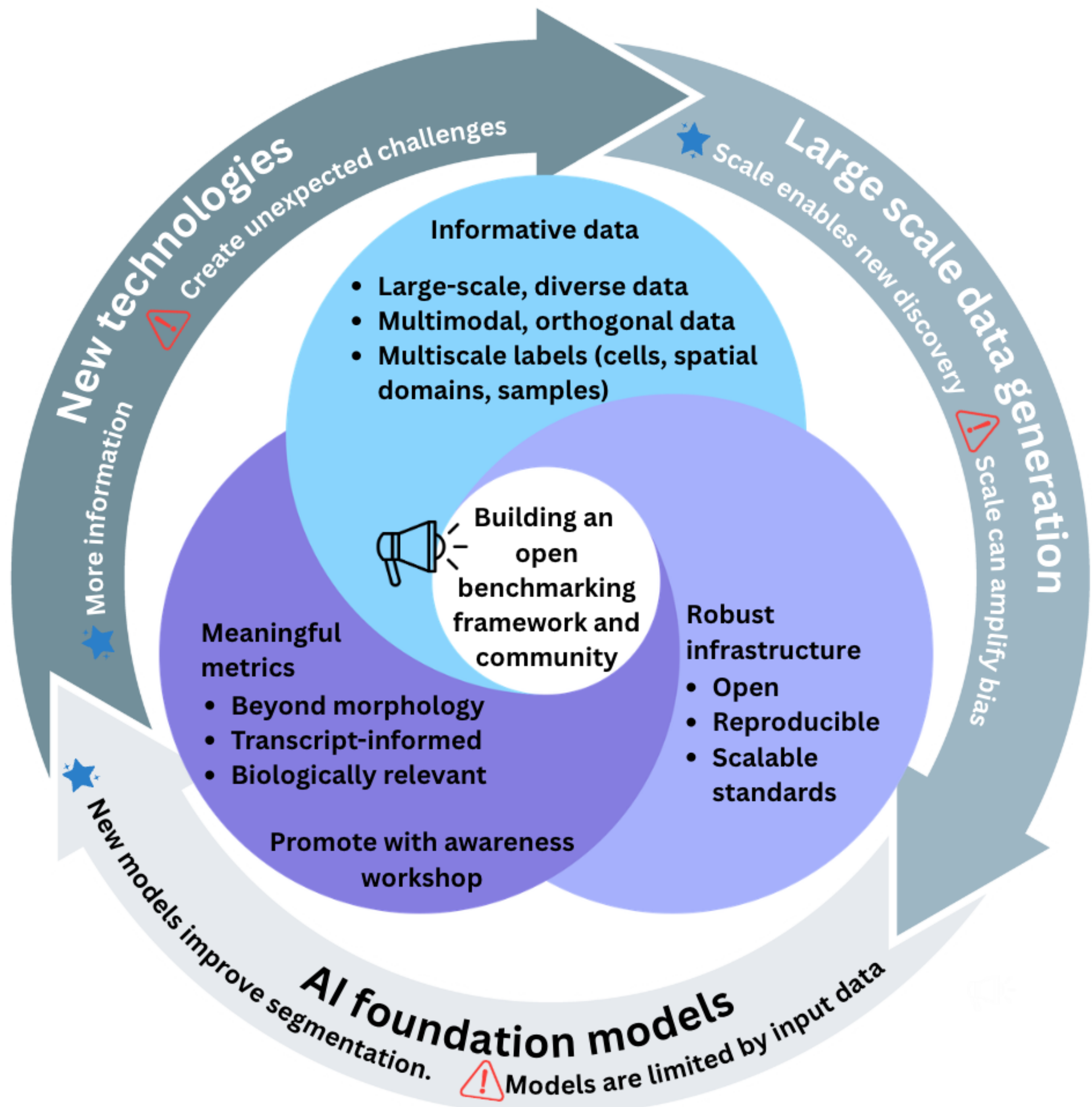

**Figure 2.** Schematic highlighting the need for a coordinated community effort to advance cell segmentation in the rapidly evolving field of computational and experimental spatial biology. At the core is the development of an open benchmarking framework and community, supported by three key pillars: (i) informative data (large-scale, multimodal, and multiscale annotations), (ii) robust infrastructure (open, reproducible, and scalable standards), and (iii) meaningful metrics (beyond morphology, transcript-informed, and biologically relevant). The outer cycle illustrates a feedback loop in which advances in data generation and technologies enable new discoveries but may also introduce or amplify biases, while improved models and segmentation methods, including AI foundation models, in turn drive further innovation and data generation.

## Box 1 | What does “cell segmentation” mean in spatially resolved transcriptomics?

Cell segmentation in SRT refers to the computational process of connecting spatial molecular measurements to cellular entities. Unlike conventional image-based segmentation, this task is not limited to delineating cell boundaries: it also involves assigning transcripts to their cells of origin and, in some cases, jointly inferring boundaries and transcript assignments to derive cell-level expression profiles. Thus, segmentation in SRT should be understood as both a morphological and molecular inference problem.

Cell segmentation of SRT data generally takes as inputs:

1. Transcript information (*i.e.* the identity and positions of the detected transcript molecules, either as individual molecule coordinates in imaging-based SRT platforms or as spatially indexed count measurements in sequencing-based platforms).
2. Sample images (if available). These often include stainings for nuclei (*e.g.* DAPI), outer cell membranes, protein stains, as well as polychromatic stains such as hematoxylin and eosin (H&E). Note that images associated with SRT technology may be collected purely for quality control (e.g. to verify tissue sample placement on a slide) and may not be suitable (e.g. may be too low resolution or unfocused) for image analysis.

The outputs of the segmentation procedure typically yield one or more of the following:

1. Instance segmentation masks of cell boundaries (in 2D or 3D pixel/voxel space)
2. Assignment of detected transcripts to cellular entities, which may be represented either as discrete assignments or as probabilities.

Cell segmentation methods for SRT can be broadly categorized into (1) “transcript-agnostic” methods, which operate on image data and aim to first accurately delineate cell boundaries before using them to assign transcripts, and (2) “transcript-informed” methods, which may be multi-modal and use transcript and potentially image data to infer transcript-to-cell assignments.

Below we describe these approaches and their respective objectives and limitations.

**Transcript-agnostic (image-based) segmentation**

Transcript-agnostic segmentation methods (e.g., StarDist (Stevens *et al.*, 2022), Cellpose (Stringer *et al.*, 2021)) perform instance segmentation in image space, using continuous morphological features extracted from nuclear, cytoplasmic and/or membrane staining to assign pixels to distinct cell masks. The output is an assignment of each pixel $pixels$ to a cell,

$$\text{pixels} \mapsto \text{cell id}$$

These methods aim to delineate cell boundaries and hence enable downstream morphometric analysis. These methods do not use transcript information (coordinates or gene identity). Instead, transcripts are subsequently assigned to the cell mask within which they spatially reside.

<u>Limitations</u>

Because of its widespread availability and ease of access, nuclear staining is often used as a proxy for whole-cell segmentation, but this typically leads to under-segmentation and exclusion of cytoplasmic regions. Nuclear expansion - radial dilation of nuclear masks (sets of pixels) to approximate cell boundaries - can partially address this limitation; however, overall cell morphology does not necessarily mirror nuclear shape. In addition, membrane or nuclear staining may be incomplete or entirely absent, further compromising boundary accuracy. For example, DAPI-based segmentation of nuclei could over-segment multinucleated cells (e.g., syncytiotrophoblasts, skeletal myofibers) and miss anucleated cells (e.g., erythrocytes) or 2D plane regions through non-nuclear cytoplasmic cell regions. Most transcript-agnostic segmentation methods operate in 2D, despite transcripts originating from multiple focal z-planes in imaging-based SRT, and measured transcript positions may be displaced outside the cell membrane due to technical limitations (**Box 2**) (Linares *et al.*, 2023; Anacleto *et al.*, 2026). Consequently, assigning transcripts solely by overlap with inferred 2D masks produces contaminated and/or incomplete cell-by-gene count matrices.

**Transcript-agnostic segmentation (pixel-based)**

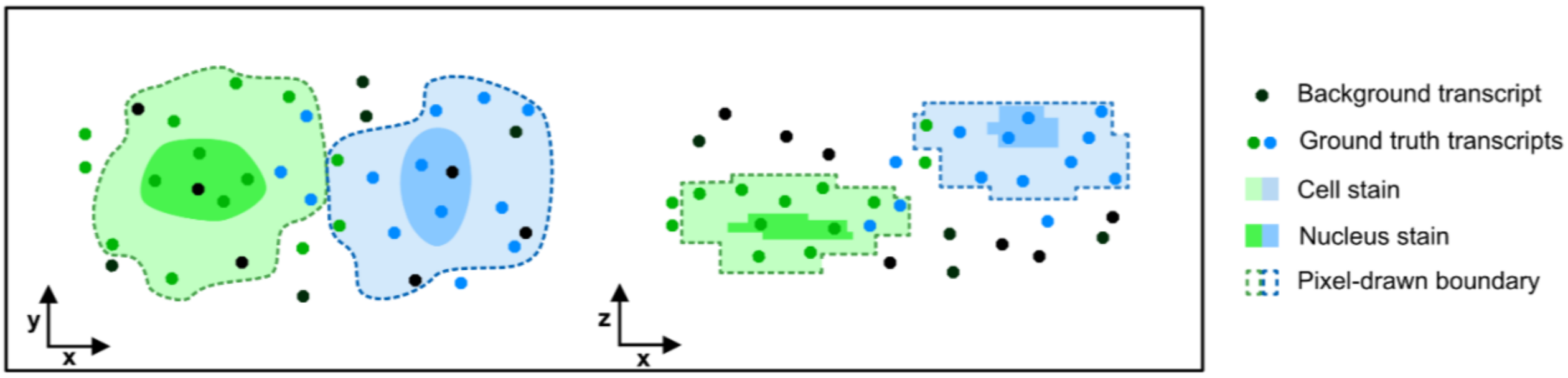

**Figure caption.** Schematic of pixel-based, transcript-agnostic segmentation. **Left (xy-plane)**: in-plane stain intensity is sampled near-continuously, so the illustration depicts an idealized case in which the segmentation boundary aligns tightly with the cell and nucleus stains. **Right (xz-plane)**: most imaging-based SRT platforms acquire z-information discretely, in stacks spaced 0.5–1.5 µm apart, so both the stain and the resulting pixel-drawn boundary appear stack-like along z. Dots indicate transcripts: green/blue mark ground-truth assignments to the two cells, black marks background.

### Transcript-informed segmentation

Transcript-informed segmentation methods (e.g. Baysor (Petukhov *et al.*, 2022), Segger (Heidari *et al.*, 2025), ProSeg (Jones *et al.*, 2025)) infer transcript-to-cell-of-origin relationships using transcript coordinates (in 2D or 3D) and gene identities (or lattice-based count fields from sequencing-based platforms), optionally incorporating priors on cell shape and gene expression. The output is an assignment of each transcript $t_j$ to a cell,

$$t_j \;\mapsto\; \text{cell id}$$

often expressed probabilistically as $p(\text{cell} = k \mid t_j)$. This directly defines the quantity used in downstream analysis, i.e. cell-level gene-expression counts, and allows reassignment of transcripts outside inferred boundaries or separation of transcripts originating from overlapping cells in 3D tissue.

<u>Limitations</u>

Transcript-informed segmentation methods are still relatively new and remain poorly characterized. Approaches that optimize transcriptional coherence may over-segment cells exhibiting subcellular transcript localization, or conversely under-segment densely packed regions composed of transcriptionally similar cell types (e.g., hepatocytes in the liver). Thus, cell masks derived directly from transcript assignments (e.g. via convex hulls) often poorly approximate true cell morphology and may not align with cell structure (e.g. membrane) staining measurements. In addition, transcript-informed methods can be sensitive to panel design, transcript detection efficiency, local RNA density, and spatial artifacts such as diffusion, spillover, or background RNA.

**Transcript-informed segmentation (transcript-based)**

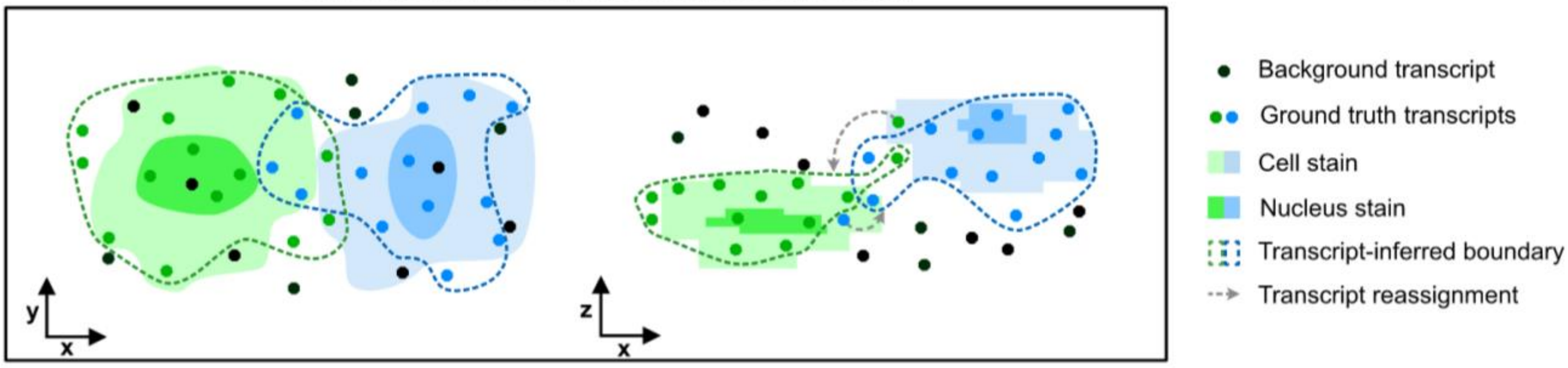

**Figure caption.** Schematic of transcript-informed segmentation; stains and transcript symbols follow Box Figure 1. **Left (xy-plane):** the illustration depicts an idealized case in which every ground-truth transcript is assigned to its correct cell, producing irregular and overlapping boundaries that bend outward to enclose distant transcripts. **Right (xz-plane):** the same irregularity carries into z, since boundaries are driven by transcript positions rather than stain. When a transcript lies too far from its plausible parent cell, some algorithms instead reassign it to a cell without redrawing the boundary; such reassignments are shown as grey arrows.

### Multimodal approaches

Most transcript-informed methods still rely on imaging modalities by conditioning transcript assignment on image-derived cell masks. More advanced approaches, such as BIDCell (Fu *et al.*, 2024) and ProSeg (Jones *et al.*, 2025), perform true multimodal inference by jointly learning cell boundaries and transcript assignments from both imaging and transcriptomic information. Throughout this work, we use the term segmentation broadly to encompass both transcript-agnostic and transcript-informed approaches, and we explicitly distinguish between them when relevant.

## Box 2 | Glossary of spatial displacement and mixing artifacts in spatial transcriptomics

To clarify literature usage of partially overlapping terminology describing how transcripts become spatially displaced and how mixed signals arise in SRT experiments, we distinguish between **(i) primary causes** and **(ii) downstream manifestations** of transcript mislocalization or mixing.

### Primary causes

These describe physical, optical, geometric, biological, or technical processes that alter, obscure, or complicate transcript localization relative to their actual positions and assumed cell boundaries prior to downstream computational analysis.

**Transcript diffusion** refers to the physical displacement of RNA molecules caused by tissue permeabilization, fixation, or related sample-processing steps, leading to lateral spread beyond original cell borders (Ståhl *et al.*, 2016; You *et al.*, 2024; Marco Salas *et al.*, 2025).

**Optical and amplification artifacts** arise in imaging-based SRT, where dense transcript signals, diffraction-limited spot overlap, autofluorescence, crosstalk, photobleaching, stage drift, amplification-related crowding, and positional uncertainty can reduce localization accuracy (Linares *et al.*, 2023; Fortner and Bucur, 2024).

**Structural and geometric assignment ambiguity** includes cases in which transcripts are not physically displaced but become difficult to assign correctly because tissue geometry is not well captured by 2D segmentation. This includes **3D overlap and projection effects**, where vertically stacked cells project into the same imaging plane, and **cell protrusions or complex morphology**, where transcripts extend beyond compact cell-body assumptions (Cajigas *et al.*, 2012; Marco Salas *et al.*, 2025; Tiesmeyer *et al.*, 2026).

**Boundary detection artifacts** arise when membrane or nuclear stains are weak or incomplete, for example, due to poor antigenicity, fixation effects, or low signal-to-noise ratio. These issues impair segmentation and can lead to missing, merged, or fragmented cell masks, causing transcript misassignment even when transcript localization itself is accurate. This is particularly pronounced for segmentation approaches that rely predominantly on morphological stains, rather than transcript information (Wu, Beechem and Danaher, 2025).

**Background contamination** refers to non-cell-bound transcripts in the local environment, including **ambient RNA** released by lysis or incomplete clearing and potentially **extra-cellular RNA** such as RNA carried in extracellular vesicles. In practice, these signals are difficult to distinguish from other unassigned transcripts and are often analyzed using approaches adapted from single-cell RNA-seq (Fleming *et al.*, 2023; Marco Salas *et al.*, 2025).

### Downstream manifestations

The causes described above lead to related manifestations of transcript mislocalization or mixing in measured cell-level data and inferred expression profiles. In sequencing-based SRT, this is often referred to as **spatial bleeding** or **spot swapping**, whereas in imaging-based SRT it is more commonly described as **spillover** (Ni *et al.*, 2022; Mason *et al.*, 2024; Crowell *et al.*, 2025; Sun *et al.*, 2025). At the cell level, the resulting mixed expression profile is typically described as **contamination** or **admixture**, i.e. the presence of transcripts from neighboring sources (Mitchel, Gao, *et al.*, 2025).

## Table 1. Impact of segmentation errors on downstream analysis.

| Analysis task & description | Mechanism of error | Severity of impact |
|---|---|---|
| **Cell type classification.** Assigning cell identities to segmented cells based on canonical marker genes or through label transfer from annotated scRNA-seq reference datasets. | Segmentation errors may mis-attribute molecules of key marker genes to the wrong cells, confusing cell-type calling algorithms. High admixture fractions create avoidable segmentation-induced doublets, diluting cell expression profiles and complicating annotation and matching with other data. | **Low/Moderate.** Algorithms use global expression signatures and are relatively robust to small perturbations in gene expression profiles. In contrast, rare cell populations (especially those that express marker genes from multiple cell types, e.g., polyhormonal cells), transitional states, or low-abundance cells are likely to be more sensitive to such discrepancies, which can impact downstream analyses. |
| **Dimension reduction and unsupervised clustering.** Grouping cells based on transcriptional similarity to discover cell states or populations without prior labels. | Contamination both blurs genuine cell-specific expression profiles and generates artificial hybrid profiles. In low-dimensional embedding spaces (e.g., PCA, UMAP), these distortions can produce spurious clusters while compressing separation among true cell populations, degrading clustering resolution. | **Moderate/High.** Clustering is highly sensitive to systematic contamination/admixture, which both generates artificial hybrid profiles and blurs genuine transcriptional differences, leading to collapsed or poorly resolved cell states. |
| **Context-dependent differential expression (DE).** Comparing the gene expression of a specific cell type across distinct spatial regions, such as tumor vs. adjacent normal or crypt vs. villus, to identify location-specific state changes. | Spatial regions vary in cell composition. This will lead to systematic bias in admixture profiles, which will show up as false positive DE results. A cell type will preferentially acquire admixture from the dominant neighbors in its specific region. For example, T cells in tumor regions will appear to upregulate malignant genes simply because they are surrounded by malignant cells. | **Very high.** Admixture artifacts frequently dominate the results and can be difficult to distinguish from true biological effects. |
| **Niche detection / spatial domains.** Identifying recurrent signal neighborhoods, tissue regions, or multicellular microenvironments based on cell composition and/or gene expression. | Segmentation errors can alter local cell-type composition and create artificial hybrid profiles, causing niches to appear more mixed, less distinct, or enriched for signals derived from neighboring cells rather than true local biology. | **High.** Niche definitions are sensitive to systematic contamination and local misassignment, especially when spatial domains are inferred from both cell identity and expression state. |
| **Neighbor cell interactions.**<br>Analyzing how a cell's gene expression depends on the presence of specific neighbors, and modeling the association between a cell's transcriptional state and the composition of its local neighborhood. | Systematic admixture appears predictive: When comparing cells "near" vs. "not near" a specific neighbor, the "upregulated" genes are often just markers of that neighbor leaking into the focal cell. Predictive models achieve high accuracy by learning to detect these leaked foreign transcripts rather than true regulatory responses. | **Very high.** The admixture from the neighbor is directly interpreted as an interaction response (e.g., malignant cell markers in T lymphocytes near tumor cells). In predictive models, the top informative genes are predominantly foreign markers. |
| **Ligand-receptor (LR) interaction inference.**<br>Prioritizing potential intercellular communication channels by identifying cognate ligand-receptor pairs expressed in spatially adjacent cell types. | Misassignment of ligand or receptor transcripts to distinct adjacent cells can create potential false-positive LR channels. Inference methods commonly look for coordinated behavior between communicating cells, and the systematic appearance of the erroneous LR pairs can score highly in that regard. | **High.** Top predicted LR interactions often involve LR pairs that are expressed in one cell type. |
| **Multicellular programs.**<br>Using factor analysis tools (e.g., DIALOGUE (Jerby-Arnon and Regev, 2022), scITD(Mitchel, Gordon, *et al.*, 2025)) to identify multicellular patterns of transcriptional state co-variation across cell neighborhoods. | The resulting factors (or programs) often capture the correlation between a cell type and its contamination source rather than biological coordination. For example, a T lymphocyte program might simply track the level of fibroblast contamination across different niches. | **Moderate.** Admixture effects are mixed with the factors, confounding biological interpretation. |

**Table 1.** We evaluate how segmentation errors affect common spatial omics analysis tasks. While some analyses (e.g., cell-type classification) are robust, high-dimensional and context-dependent analyses (e.g. clustering, spatial differential expression, neighbor cell interaction modeling, and ligand–receptor inference) are more sensitive to systematic bias, often leading to spurious cell states, false interactions, and misleading biological conclusions.